\documentclass[11pt]{fdmp}
\usepackage{graphicx}
\usepackage{subfigure}

\def\egbert#1{} 

\begin{document}
\fdmp{1}{1}{2999}{100}

\runningtitle{Numerical study of liquid metal flow}

\title{Numerical study of liquid metal flow in a rectangular duct \\
 under the influence of a heterogenous magnetic field}

\author{
    Evgeny~V.~Votyakov\thanks{Institut f\"ur Physik,
     Technische Universit\"at Ilmenau, PF 100565, 98684 Ilmenau, Germany},
    Egbert A.~Zienicke
    }

\maketitle

\abstract{We simulated numerically the laminar flow in the
geometry and the magnetic field of the experimental channel used
in [\cite{Andrejew:Kolesnikov:2006}]. This provides detailed
information about the electric potential distribution for the
laminar regime (numerical simulation) and in the turbulent regime
as well (experiment). As follows from comparison of simulated and
experimental results, the flow under the magnet is determined by
the interaction parameter $N=Ha^2/Re$ representing the ratio
between magnetic force, determined by the Hartmann number $Ha$,
and inertial force, determined by the Reynolds number $Re$. We
compared two variants: (i) $(Re,N)$=(2000,18.6)\ (experiment),\
(400,20.25) (simulation), and (ii) $(Re,N)=$(4000,9.3)
(experiment), (400,9) (simulation) and found an excellent
agreement for the numerical and experimental distributions of the
electric potential. This is true despite of the fact that the
experimental inflow is turbulent while that in the simulation is
laminar. As a special feature of the electric potential
distribution local extrema under the magnets are observed, as well
experimentally as numerically. They are shown to vanish, if the
interaction parameter falls below a critical value. Another
interesting new detail found in our numerical calculations is the
appearance of helical paths of the electric current. Using a
simplified magnetic field without span-wise dependence, we show
that important physical features of the considered problem are
sensitive to small variations in the spatial structure of the
magnetic field: the local extrema of the electric potential and
also the helical current paths disappear when the simplified
magnetic field is used. The structure of the three dimensional
velocity field is also investigated, in particular, a swirling
flow is found in the corners of the duct caused by Hartmann layer
destruction behind the magnets.}


\section{Introduction}

The flow of an electrically conducting fluid under a localized
inhomogeneous magnetic field is of interest for many industrial
applications dealing with the problem to influence hot metal melts
by the use of magnetic fields. This has the advantage that no
direct contact with the chemically aggressive melt is necessary
[\cite{Davidson:Review:1999}]. One prominent example is the
electromagnetic brake used in modern continuous steel casting,
see for instance   
[\cite{Takeuchi:Proc:2003}]. If one neglects the liquid steel jets
entering the mould, the flow in a liquid metal channel under a
static localized magnetic field may reproduce qualitatively many
features of the electromagnetic brake: the braking effect on
stream-wise velocity, suppression of turbulence under the magnet,
and effects of strong spatial dependence of the magnetic field.
Another example, where liquid metal flow in a channel is important
for possible industrial applications, is the Lorentz Force Velocimetry
based on exposing the fluid to a magnetic field and measuring the
drag force acting upon the magnetic field lines
[\cite{Thess:Votyakov:Kolesnikov:2006}].

In the recently appeared experimental work
[\cite{Andrejew:Kolesnikov:2006}] the liquid metal flow was
systematically investigated for the range of interaction parameters
$4 \leq N \leq 20$. In the experiments special attention was
focused on the suppression of turbulence by the magnetic field and
on a systematic recording of data for the electric potential
building up under and around the magnets. As a result, the full
map of the electric potential distribution was obtained for the
middle horizontal plane of the rectangular duct and few $N$
parameters. If the transverse magnetic field would be homogenous
the electric potential data might be used to determine the
velocity components perpendicular to the magnetic field
[\cite{Sommeria:Moreau:MHDturblulence:1982}]. In the present case of
strongly inhomogeneous magnetic field, the electric potential data
can be solely compared with the results of the corresponding
numerical simulations.

\begin{figure}[tb]
\begin{center}
    \includegraphics[angle=0, width=3.5in, clip]{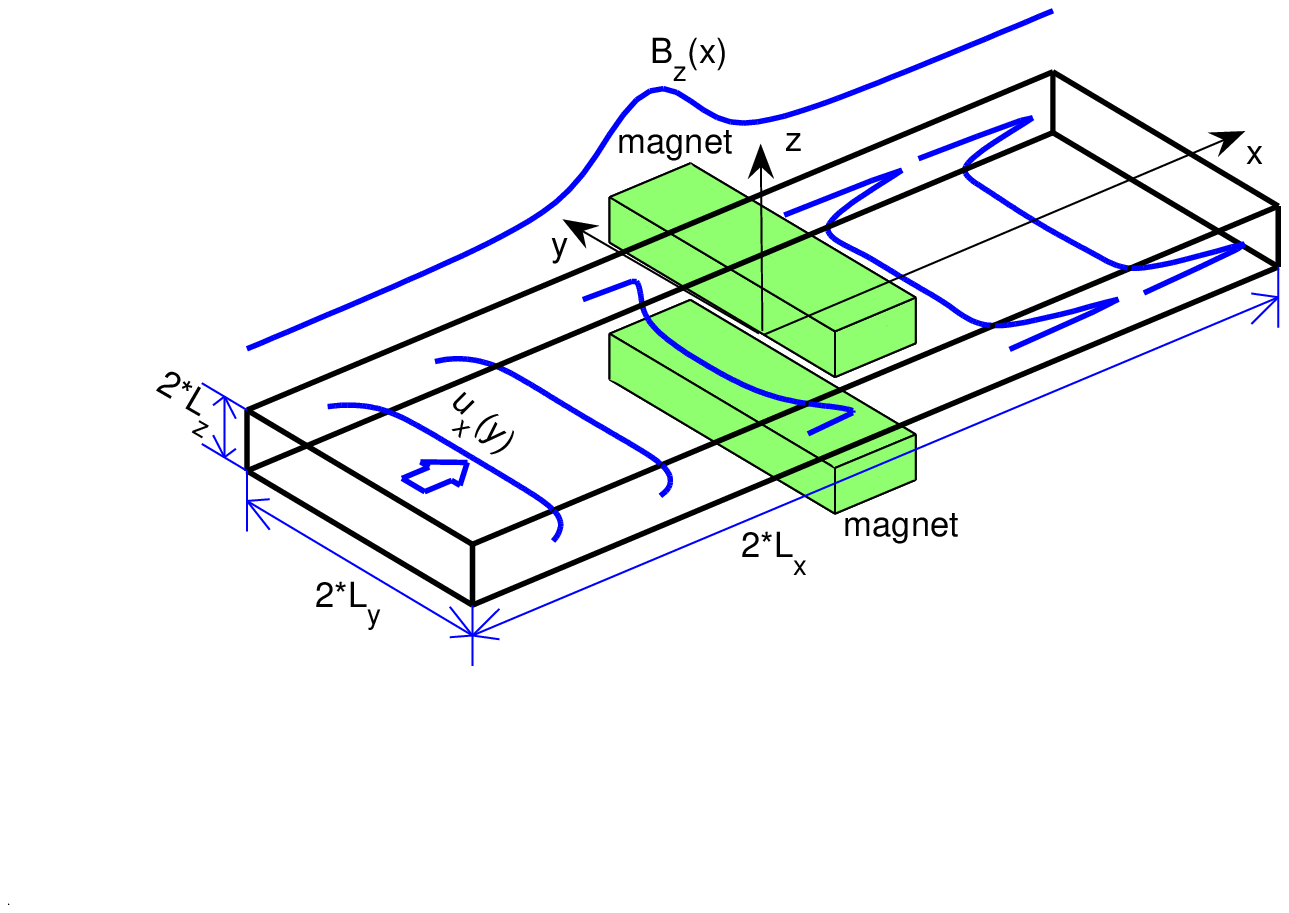}
  \caption{Coordinate system and sketch of the problem: rectangular channel
  $2L_x\times 2L_y\times 2L_z$ and
  two magnets on the bottom and top walls. Qualitatively,
  there are shown transverse magnetic field $B_z(x)$ varying along stream-wise
  direction, and span-wise profile of stream-wise velocity $u_x(y)$.
  The center of the coordinate system is in the center of the magnetic gap.}
  \label{Fig:ChannelMagnet}
\end{center}
\end{figure}

Fig.~\ref{Fig:ChannelMagnet} presents schematically the geometry
and magnetic field configuration studied in
[\cite{Andrejew:Kolesnikov:2006}]: the liquid metal moves in a
rectangular duct, where the locally heterogenous magnetic field is
created by two permanent  magnets on the top and bottom walls of
the duct. The originally convex velocity profile $u_y(x)$ adopts,
by passing the magnetic field $B_z(x)$, a characteristic M shape
what is a manifestation of the electromagnetic brake process.

The main goal behind our numerical investigation is first to
reproduce features of the flow and the electromagnetic quantities
found in the experiment. Moreover, since numerical calculations
have the advantage that all quantities are available in the whole
computational domain, the second goal is to visualize the
additional data, i.e. 3D velocity field and electric current
paths,  which are not accessible to measurements in the
experiment. Before we explain why one can successfully
compute the experimental flow under and near the magnet
using laminar calculations let us shortly describe in the next
paragraphs the state of the art of numerical and theoretical
approaches that could be appropriate to describe the present
problem involving turbulence and inhomogeneous magnetic field at
the same time.

The most ideal approach from the point of view of physical exactness
would be to carry out a direct numerical simulation (DNS) of the
experimental flow which models correctly the whole flow including
turbulent regions down to small scales.  However, to do this
under a magnetic field steeply varying with space coordinates
is extremely hard since it requires space and time resolution a few
orders of magnitude more than is available today. Actually, to
catch all spatial structures of the flow, especially Hartmann
layers and sidewall jets, the simulation has to be fully
three-dimensional and needs high resolution near the boundaries.
This concerns also the turbulent flow which was generated in the
experiment by a honeycomb at the inflow in order to observe suppression
of fluctuations by the static magnetic field. The inflow distance
takes almost half the length of the computational domain, therefore
half of the computational resources must be paid to simulate
correctly the flow without explicit action of the magnetic field,
just to monitor declining turbulent fluctuations.

Other numerical approaches to catch turbulent features of the flow
are Large eddy simulations (LES) and Reynolds averaged stress models
(RANS). These are as well faced with serious
technical difficulties. The present state of the art for LES
of conducting fluids under magnetic fields is such that one may
treat the flow inside a homogenous magnetic field only if one
resolves the boundary layers as in a direct numerical simulation
(DNS), see [\cite{Knaepen:Moin:2004,Vorobev:etal:2005}]. The
definition of appropriate wall functions for the turbulent
Hartmann and sidewall layers is still an actual field of research.
LES in inhomogeneous magnetic field meets serious difficulties as
one has to find appropriate methods of spatial averaging. For
the momentary available RANS models it is not yet clear whether
they are able to describe the unavoidable anisotropy of the
turbulent scales inside strongly varying magnetic field, see
[\cite{Widlund:Zahrai:Bark:1998,Kenjerec:Hanjalic:2000,Kenjerec:Hanjalic:2004}].
In any case, before to start a LES or RANS study one first has to
define and verify parameters appearing in these phenomenological
models by DNS runs.

The usual analytical means also fail to describe the features of
the experimental flow. This holds, because the parameters of
the applied magnetic field lie outside the limits imposed by the
 assumptions necessary for any analytical theoretical
treatment. In particular, to
neglect inertial effects, the existing analytical approaches
assume very strong and slowly varying magnetic field, which is not
fulfilled neither for the electromagnetic brake nor for the
channel flow considered here. Typically, any regular theoretical
method is based on an asymptotic expansion of MHD equations around
large $N$ [see for example \cite{Lavrentiev:Molokov:Magnetohydrodynamics:1990}
and references therein].
However, in the system under consideration, the local interaction
number $N(x)=B_z(x)N$ goes up from zero to moderately high
values on a short distance under the inward gradient of the
magnetic field. Moreover, even if such an approach is not entirely
impossible for some cases, it employs a series expansion what is
of the same cost approximately as a full 3D simulation
[\cite{Sellers:Walker:1999}]. Also, the theoretical methods do not
take a span-wise dependence of the magnetic field into account,
however, as we shall see later, this seemingly fine detail of the
magnetic field configuration can be responsible for significant
qualitative features in the electric potential distribution inside the
magnetic gap.

However, as follows from the experimental data
[\cite{Andrejew:Kolesnikov:2006}], the intensity of turbulent
fluctuations inside the magnetic gap is lower than one percent,
and essentially smaller than at the inlet distance at front of the
magnetic system. This provides evidence that the magnetic field is
strong enough to be the main influence shaping the flow structure
inside the magnetic gap. This flow structure, qualitatively
characterized by a M-shaped  profile, is weakly dependent on the
separate $Re$ and $Ha$ numbers and strongly dependent on the
interaction parameter $N$, especially when $N$ is high. Another
conjecture from these experimental data is that the originally
turbulent inlet velocity profile is of small importance as well.

The foregoing statements give us an opportunity to reproduce the
experimental results using a laminar numerical 3D simulation. That
means, we do not carry out a computation with the same values of
$Re$ and $Ha$ as in the experiment and do not monitor turbulence.
Instead, since the main effects are due to the interaction number
$N$, one may select for the simulation $Re$ and $Ha$ lower than in
the experiment but belonging to the same ratio $N=Ha^2/Re$. As inlet
flow one takes a laminar duct flow. To clarify, whether the shape
of inflow velocity plays a role for the measured  experimental
data, one can test different laminar inflow profiles having the
same mean flow rate but different flatness.

It is easy to point out a reasonable range for the $Re$ numbers
in the numerical simulation. Large $Re$ parameter provokes
turbulence which could not be properly resolved with current
computational resources while too low $Re$ number results in a
viscous force in the core of the flow. On the other hand, the
highest  limit of the $Re$ number is governed also by the value of
$Ha$ number which is, in its own turn, dictated by the available
grid resolution for the Hartmann layer.

Thus, the main goals of the simulation were to determine the
qualitative global velocity field and to find a good reproduction
of the electric potential in the magnetic field region.
Especially, we were interested to find the two extrema of the
electric potential which were observed in the experiment. As it
turns out, all these aims were reachable. The overall general
features of the laminar flow are well represented by our code. In
section~\ref{subsec:flow} we present how the Hartmann layer and
the sidewall jets are forming under the magnet in a stationary
flow. For the electric potential under the magnet the experimental
results of [\cite{Andrejew:Kolesnikov:2006}] show clearly
(section~\ref{subsec:epot}) that the flow under the magnet is near
to being laminar. In this region the magnetic and the inertial
forces are predominant, if one excludes the regions very near to
the walls where the viscous forces are essential. Therefore we
find a good representation of the electric potential distribution
of the experiment for our runs using lower Hartmann and Reynolds
numbers but keeping the same interaction parameter.

The structure of the present paper is the following. In section
\ref{sec:eq} the equations and our numerical method to solve them
are presented. Here also the inflow profiles and the used grid are
specified. In section \ref{sec:mfield} some properties of the
experimental
magnetic field are explained and a second simplified magnetic
field with no span-wise dependence is introduced, which serves to
show that small changes on the inhomogeneous magnetic field can
lead to remarkable differences in the electric potential
distribution. In section \ref{sec:results} we present the results
of our numerical computations \egbert{begin added} showing all
characteristics of the velocity field (section \ref{subsec:flow}),
the comparison of the numerically determined electric potential
with experimental data (section \ref{subsec:epot}), and the
electric current paths (section \ref{subsec:current}). \egbert{end
added}

\section{\label{sec:eq}Equations and numerical
            method}

\egbert{taken from below, to have explanation of quasi-static
approximation and references at once} The governing equations for
electrically conducting and incompressible fluid are derived from
the Navier-Stokes equation coupled with the Maxwell equations for
moving medium, and also using the Ohm's law. We apply the
quasi-static (induction-less) approximation where it is assumed that
an induced magnetic field is infinitely small in comparison to the
external magnetic field (see, e.g. [\cite{Roberts:1967}]), so it is
neglected when one
calculates the Lorentz force, but it is not neglected at finding
the electric current density {\bf j}. \egbert{end taken from
below} The resulting equations in dimensionless form are then
given as follows:
    \begin{eqnarray}
    \label{eq:NSE:momentum}
    \frac{{\partial\textbf{u}}}{{\partial t}} + (\textbf{u} \cdot \nabla ) \textbf{u}
    &=& - \nabla p + \frac{1}{Re}\triangle \textbf{u} + N
    (\textbf{j}\times\textbf{B}),
    \\
    \label{eq:NSE:continuity} \nabla \cdot \textbf{u} &=& 0,
    \\
    \label{eq:NSE:Ohm} \textbf{j} &=&   -\nabla\phi + \textbf{u} \times \textbf{B},
    \\
    \label{eq:NSE:Poisson} \triangle \phi &=& \nabla \cdot(\textbf{u}\times\textbf{B}).
    \end{eqnarray}
Here $\bf{u}$ denotes velocity field, $\bf{B}$ is an external
magnetic field, $\bf{j}$ is electric current density, $p$ is
pressure, $\phi$ is electric potential, $Re=u_0H/\nu$ is Reynolds
number, $N=Ha^2/Re$ is the interaction parameter (Stuart number),
and $Ha=HB_0(\sigma/\mu)^{1/2}$ is Hartmann number, all defined
with the half-height of the channel $H$, mean velocity $u_0$,
typical magnetic field strength $B_0$, density $\rho$, electric
conductivity $\sigma$, kinematic $\nu$ and dynamic $\mu=\rho\nu$
viscosities.

As follows from eq.\ (\ref{eq:NSE:momentum}) the viscous force
$\Delta {\bf u}$ is scaled by Reynolds number $Re$, therefore at
high $Re$ and far from the walls it plays a minor role. As a
result the flow is governed by the interaction parameter $N$
defining  the ratio between magnetic and inertial forces. This is
a case we treat in the paper under consideration by comparing
experimental and simulated results with similar $N$ and different
(and high) $Re$ numbers.

In the experiments [\cite{Andrejew:Kolesnikov:2006}], the eutectic
alloy $\mbox{Ga}_{0.68}\mbox{In}_{0.20}\mbox{Sn}_{0.12}$ was used
as a liquid metal. It has density $\rho= 6360 \,\ \mbox{kg/m}^3$,
electric conductivity $\sigma=3.46\cdot 10^6 \,\ \mbox{Ohm}^{-1}$
and kinematic viscosity $\nu= 3.4\cdot 10^{-7} \,\
\mbox{m}^2/\mbox{s}$. Thus, the Hartmann number
$Ha=HB_0(\sigma/\rho \nu)^{1/2}$ defined with half-height of the
channel ($H=1\,\,\mbox{cm}$) and magnetic field intensity
$B_0=0.504 \,\,\mbox{T}$ is $Ha=193$. The interaction parameter
$N=Ha^2/Re$ was varied in the experiments by means of the mean
velocity rate $u_0$ entering the Reynolds number $Re=Hu_0/\nu$.
We have implemented the experimental range of interaction parameters by
varying mainly $Ha$ (up to 120) and keeping $Re=400$, thus in the
simulation  $4 \leq N \leq 36$.

Unknowns of the equations  (\ref{eq:NSE:momentum} --
\ref{eq:NSE:Poisson}) are the velocity vector field ${\bf
u}(x,y,z)$, and two scalar fields: pressure $p(x,y,z)$ and
electric potential $\phi(x,y,z)$. The domain of the flow is given
by a rectangular channel (Fig.~\ref{Fig:ChannelMagnet}) ($|x| \leq
L_x$, $|y| \leq L_y$, and $|z| \leq L_z$ with $L_x=25$, $L_y=5$,
and $L_z=H=1$) having the same aspect ratios
($\mbox{Length:Width:Height}$=25:5:1) as the experimental channel
of [\cite{Andrejew:Kolesnikov:2006}]. (In this experimental paper,
$H=2$ is defined as a whole height of the channel, and in the
present paper we take $H=1$ as a half-height. Thus, the $Re$ and
$Ha$ numbers given in [\cite{Andrejew:Kolesnikov:2006}] were
divided by factor two here.)  For the external magnetic field we
used the magnetic field which was measured in the experiment and a
second configuration, which is used for comparison (see section
\ref{sec:mfield}). The Hartmann number is based on the $B_0$ value
in the center of the magnetic gap $x=0$, $y=0$, $z=0$. Note, this
$B_0$ value is not the maximal one, to fulfill curl- and
divergence free requirements, the field is slightly increasing by
approaching top and bottom walls at fixed $x=0$ and $y=0$.

To solve the partial differential equations (\ref{eq:NSE:momentum}
-- \ref{eq:NSE:Poisson}) initial and boundary conditions have to
be provided. Since we are interested in a stationary solution the
initial conditions play no role (except for the speed of
convergence), and for the boundary conditions we suppose a duct
with electrically insulating and "no-slip" walls on the sides, top
and bottom. Insulating and "no-slip" conditions require at the
boundary $\Gamma$ to impose that $u|_\Gamma=0$,
$\partial\phi/\partial{\bf n}|_\Gamma=0$, where ${\bf n}$ is
normal vector to $\Gamma$. The outlet was treated as a force free
(straight-out) border for the velocity. The electric potential at
inlet and outlet borders was taken equal zero because the inlet
and outlet are sufficiently far from the region of magnetic field.

For the inlet profile we made use of two possibilities. A
self-consistent choice for laminar flow is the stationary laminar
profile of an infinite rectangular duct known analytically in the
form of a series expansion. In the experimental channel, however,
the inflow is generated by a honeycomb shaping a more flat inflow
profile in span-wise and vertical direction and generating vortex
structures which give rise to decaying turbulence on the way of
the liquid metal to the magnet. To study the influence of a more
flat turbulent like profile on the electric potential under the
magnet, we have generated a second inlet profile in the following
well defined way: periodic boundary conditions (outflow=inflow)
were imposed in stream-wise direction and then turbulent runs for
the Reynolds number of the experimental system were performed.
Then, by a space averaging procedure we computed the turbulent mean
inflow profile. In spite of this careful check of the influence of
inlet boundary conditions, we have found no difference in the
final results between the laminar and the more flat turbulent
inflows, except for the transitional region before the magnet.
This shows that the magnetic field is sufficiently strong to
completely govern the hydrodynamics of the flow when the magnetic
region is reached. It agrees also with the experimental
observation about measured turbulent fluctuations essentially
suppressed inside the magnetic gap.

\begin{figure}[t]
\begin{center}
    \includegraphics[width=3.3in]{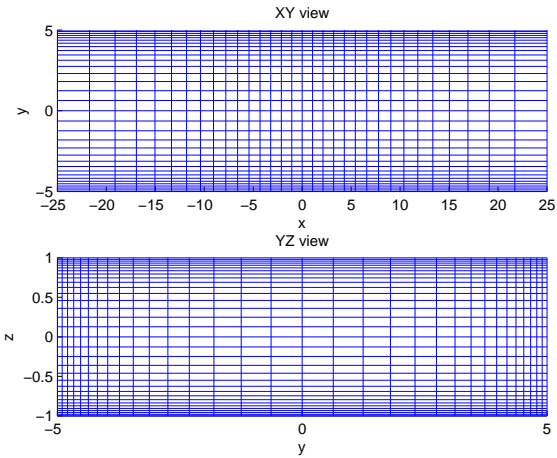}
\end{center}
    \caption{$XY$ and $YZ$ projections of the grid for
    the partition $N_x \times N_y \times N_z=32\times 32\times 32$ and $R=1.25$,
    $S=0.75$, $T=0.75$.}
    \label{Fig:Grid}
\end{figure}

As a base for our solver we used  NaSt3DGP - the simulation code
developed in [\cite{Griebel:book:1995})]. Originally, this finite
difference solver was designed for pure hydrodynamical problems.
Therefore we had to extend it by the following features to be able
to solve MHD problems: (1) using the Poisson solver also for the
determination of the electric potential, and (2) including the
Lorentz force contribution on the right hand side computing the
preliminary velocity field. Moreover, we reorganized the input and
output parts of NaSt3DGP in order to work with the arrays keeping
magnetic field, electric potential, and electric current.

Briefly, the numerical algorithm is the following. To decouple
Eq.~(\ref{eq:NSE:momentum}) - (\ref{eq:NSE:Poisson}), the
Chorin-type projection method is applied
[\cite{Peyret:Taylor:1983, Hirsch:1988}]. This is a general
procedure based on a predictor-corrector approach. First, the
Poisson equation for the electric potential (\ref{eq:NSE:Poisson})
is solved and the electric current is found according to
(\ref{eq:NSE:Ohm}). Next, a preliminary velocity field is computed
from the momentum equation without regarding the influence of the
pressure term. The second part of the time step from the
preliminary velocity field, which is not divergence-free, to the
new divergence-free velocity field allows to derive a Poisson
equation for the new pressure. Thus, the whole algorithm is
written down  as follows (index $n$ denotes time integration step,
the spatial discretization is omitted):

\begin{figure}[tb]
\begin{center}
    \includegraphics[angle=0, width=3.0in, clip]{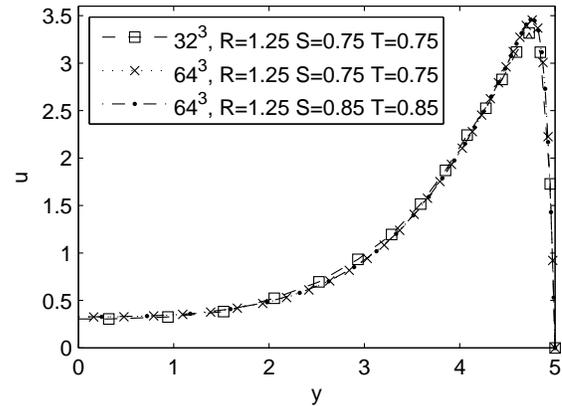}
  \caption{span-wise profile of the stream-wise velocity computed at different grid parameters:
  $N_x \times N_y \times N_z =32 \times 32 \times 32$ (squares) and $64 \times 64 \times 64$ (crosses and dots),
  $(R,S,T)=(1.25, 0.75, 0.75)$ (squares and crosses) and $(1.25, 0.85, 0.85)$ (dots).
  Other parameters are common: $Re=100$, $N=16$, $x=0$, $z=0$.}
  \label{Fig:GridComparison}
\end{center}
\end{figure}

\begin{figure*}[tb]
\begin{center}
\includegraphics[width=6.8in]{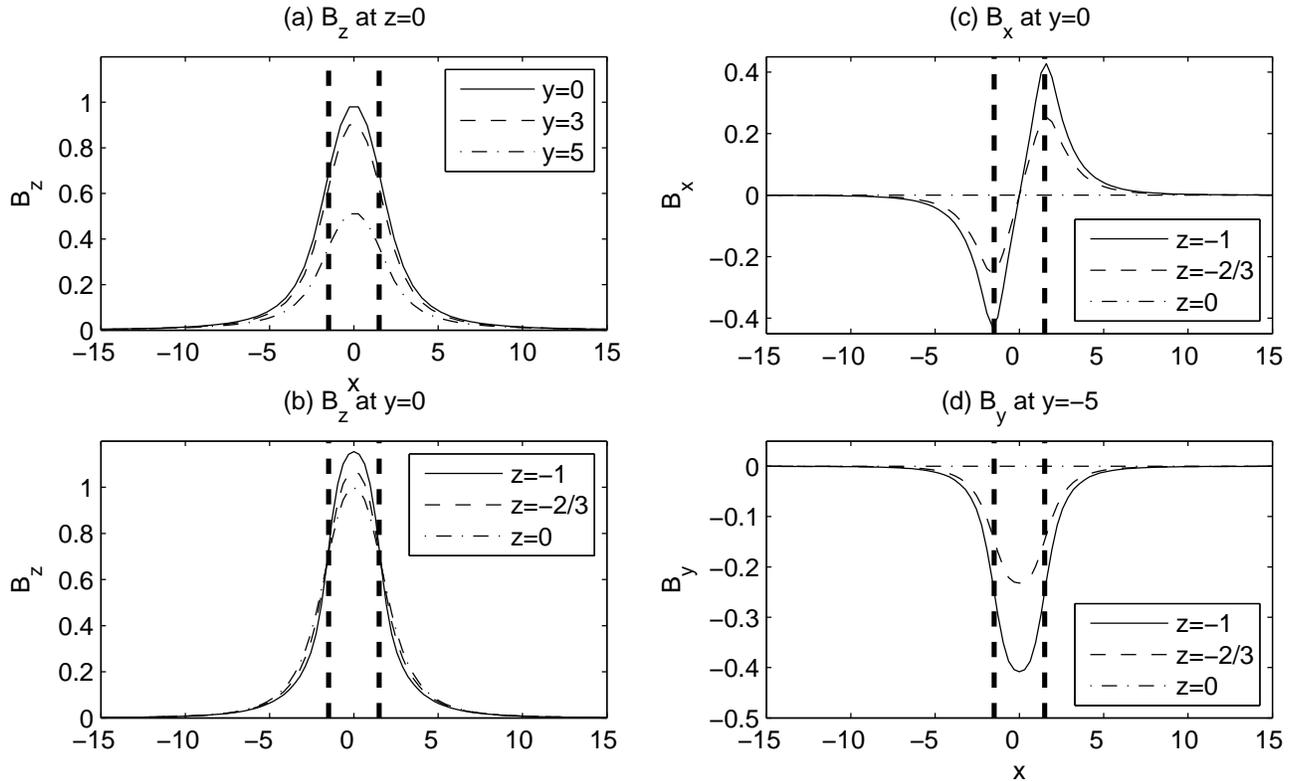}
    \caption{Cuts of the experimental magnetic field: transverse
    (plot $a$ at $z=0$ and plot $b$ at $y=0$),
    stream-wise (plot $c$ at $y=0$) and span-wise (plot $d$ at $y=-5$) components.
    For plot $a$: $y=0$ (solid), $y=3$ (dashed), and $y=5$ (dot-dashed);
    for plots $b-d$:  $z=-1$ (solid), -0.66 (dashed), and 0 (dot-dashed).
    Plots $b$ and $c$ corresponds also to the simplified spanwise-homogeneous
    magnetic field discussed in the text. Bold dashed lines are
    the borders of the physical magnet.}
    \label{Fig:ProfilesMagneticField}
\end{center}
\end{figure*}

\begin{enumerate}

\item Solve Poisson equation for the electric potential:
$$
    \triangle \phi^n = \nabla \cdot ( {\bf u}^n\times{\bf B}).
$$

\item Compute electric current:
$$
    {\bf j}^{n} =  -\nabla\phi^n + {\bf u}^n\times{\bf B}.
$$

\item With the known  ${\bf u}^n$ and ${\bf j}^n$, find the
preliminary velocity field $\tilde{{\bf u}}$
$$
    \frac{{\tilde{{\bf u}}} - {\bf u}^n}{\delta t} =
    \frac{1}{Re}\triangle {\bf u}^n + N({\bf j}^n\times{\bf B}) -
    ({\bf u}^n \cdot \nabla ){\bf u}^n.
$$

\item To compute the velocity field ${\bf u}^{n+1}$ of the next
integration step one has to solve the Poisson equation for pressure
$$
\triangle p^{n+1}=\frac{1}{\delta t} \nabla \cdot \tilde{{\bf u}},
$$
and, as a result one finds
$$
{\bf u}^{n+1} = \tilde{{\bf u}} - \delta t \nabla p^{n+1}.
$$
\end{enumerate}

The above algorithm is explicit and for simplicity it is presented
as a scheme of first order precision. It describes the principal
sequence, which in the code is realized with the Adams-Bashforth
time step having second order precision. For pressure
stabilization we apply a staggered grid, and the VONOS
(variable-order non-oscillatory scheme) scheme is used to
discretize the convective and diffusive terms.

The solver is implemented to support parallel computation: the
channel in the program code is subdivided into domains, and each
domain is run as a separate process. Communication between the
processes takes place at every integration step to synchronize
the borders between the domains.

To make sure that all the layers are properly resolved in the
simulation, we use an inhomogeneous 3D grid constructed in the
following way. First, we map $-L_x \leq x \leq L_x$, $-L_y \leq
y\leq L_y$, $-L_z \leq z \leq L_z$, onto three auxiliary variables
$-1 \leq r \leq 1$,$-1 \leq s \leq 1$, $-1 \leq t \leq 1$ as follows:
$$
    r=\frac{th(R\frac{x}{L_x})}{th(R)}, \,\,\,
    s=\frac{tg(S\frac{\pi y}{2L_y})}{tg(S\pi/2)}, \,\,\,
    t=\frac{tg(T\frac{\pi z}{2L_z})}{tg(T\pi/2)},
$$
and then variables $r,s,t$ are uniformly partitioned into $N_x$,
$N_y$, $N_z$ parts. There are three stretching parameters $R,S,T$
providing a denser ($R$) grid at $x=0$ and close to walls ($S,T$).
Typical values used in the simulation are $R=1.25$, and $0.75\leq
S \leq 0.95$, $0.75 \leq T \leq 0.95$. Fig.~\ref{Fig:Grid} gives
an example of the grid for the partition $N_x \times  N_y \times
N_z=32\times 32 \times 32$ and $R=1.25$, $S=0.75$ $T=0.75$. For
each value of the Hartmann and Reynolds number in calculations,
parameters $R,S,T$ as well as the number of grid points were
varied in order to obtain grid-independent results, see
Fig.~\ref{Fig:GridComparison}. The typical values used for the
simulation were grid $64\times 64 \times 64$ and $R=1.25$,
$S=0.85$ $T=0.85$.

\section{\label{sec:mfield}Magnetic Field}

The magnetic field used in the present paper was measured at
equidistant points in the experiments
[\cite{Andrejew:Kolesnikov:2006}], including all three components,
and then interpolated on the grid points and stored into a
three-dimensional array which is supplied as an input to the
solver. This is the most general approach which enables us to work
with any magnetic field configuration supplied externally. The
details about the field are given in the experimental paper
[\cite{Andrejew:Kolesnikov:2006}], here we just remind that it is
created by means of two permanent magnets fixed outside on the top
and bottom walls of the channel as is shown in
Fig.~\ref{Fig:ChannelMagnet}. In the gap between the magnetic
poles the field is aligned mainly along the vertical direction
parallel to the $z$-axis. Outside of the gap, before and behind
the poles, there are regions characterized by an inward and an
outward gradient of the transverse magnetic field, see
Fig.~\ref{Fig:ProfilesMagneticField}($a$, $b$), where also the
stream-wise component, $B_x$, plays a role
(Fig.~\ref{Fig:ProfilesMagneticField}($c$)). Since  the physical
magnets are finite in the span-wise direction ($y$-coordinate), the
magnetic field is also dependent on the $y$ coordinate, in
particular, on the side walls a $B_y$ component different from
zero appears, Fig.~\ref{Fig:ProfilesMagneticField}($d$). Moreover,
the vertical component $B_z$ is rather lower near the side walls
($y=\pm L_y$) in relation to $B_z$ in the center ($y=0$)
(Fig.~\ref{Fig:ProfilesMagneticField}($a$)). A few distinguished
cross-sections of transverse, stream-wise, and span-wise magnetic
field components along the $x$-axis are shown in
Fig.~\ref{Fig:ProfilesMagneticField}. Other detailed plots of the
magnetic field can be found in the experimental paper
[\cite{Andrejew:Kolesnikov:2006}].

In addition to the experimental magnetic field configuration, which
changes slightly along the $y$-axis, a few variants were simulated for
a simplified magnetic field (in the following always referred
to as 'simplified magnetic field' in contrast to the 'experimental
magnetic field') being independent of the $y$ coordinate.
The span-wise homogeneous field in this case
originates from external magnets which are infinitely long in
$y$-direction. It is characterized only by stream-wise $B_x(x,z)$
and transverse $B_z(x,z)$ components which coincide with
$B_x(x,y,z)|_{y=0}$ and $B_z(x,y,z)|_{y=0}$ in the vertical slice
($y=0$) shown in Fig.~\ref{Fig:ProfilesMagneticField}($b,c$). It
turns out that the span-wise decline of $B_z$ near the side walls
as well as that of the other $\bf B$ components is of decisive
influence on both the electric potential distribution and electric
currents under the magnetic poles (see discussion in sections
\ref{subsec:epot} and \ref{subsec:current}). This illustrates an
importance of fine details of the localized heterogenous magnetic
field, which are usually neglected in theoretical computations.

%

\section{\label{sec:results}Results of the simulation}

This section is divided in three subsections focusing on different
physical quantities --- the velocity field, the electric
potential, and the vector field of the current density --- giving
together an overview of the general and specific features of the
experimental system. The known general features are the formation
of a Hartmann profile in the transverse direction and the
formation of an M-shaped profile in span-wise direction as well in
the region of increasing and in the region of decreasing magnetic
field (counted in stream-wise direction). In the first subsection
we demonstrate that these two processes lead to a complex
three-dimensional flow structure, which in the given magnetic
field and channel proportions only can be determined numerically.
For the electric potential distribution a direct comparison with
experimental data is presented in the second subsection. A special
feature showing up is a pair of extrema of the electric potential
under the magnet. These extrema disappear when either the
interaction parameter $N$ is too small or the simplified span-wise
uniform magnetic field (at any value of $N$) is used. The same sensible
dependence of phenomena on the magnetic field structure is found
in the last subsection for complicated helical current paths,
which are present for the experimental magnetic field and are absent for
the simplified magnetic field.


\subsection{\label{subsec:flow}3D velocity field}

\begin{figure}[tb]
\begin{center}
\includegraphics[width=3.3in]{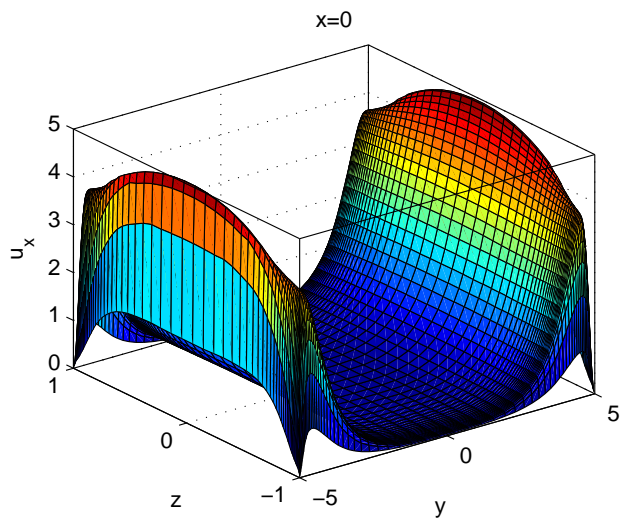}
    \caption{3D M-shaped  velocity profile, $x=0$, $Re=400$, $N=36$}
    \label{Fig:VelocitySurface}
\end{center}
\end{figure}

\begin{figure}[tb]
\begin{center}
    \includegraphics[angle=0, width=3.3in, clip]{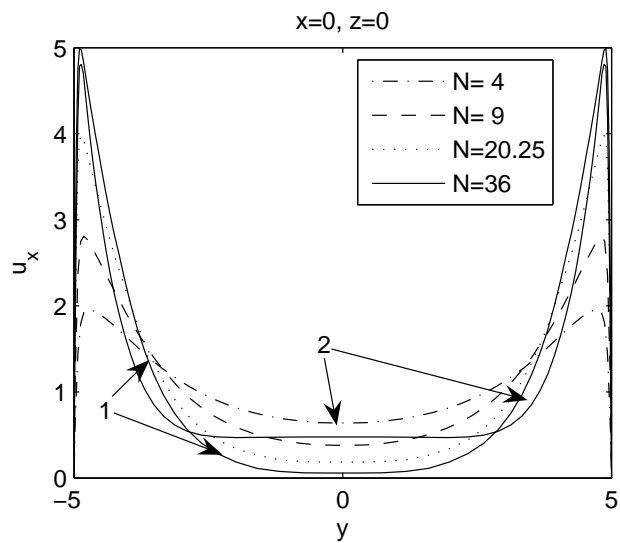}
  \caption{Stream-wise velocity profiles in span-wise direction in the center of
    the magnetic  gap ($x=0$ and $z=0$), $Re=400$ and  $N=4$ (dot-dashed),
    $N=9$ (dashed), $N=20.25$ (dotted),  $N=36$, experimental field
    (solid line 1) and  $N=36$, simplified field (solid line 2).}
  \label{Fig:VelocityProfilex0}
\end{center}
\end{figure}

We start from the well known phenomenon for duct flow under a
locally heterogenous magnetic field: the M-shaped  velocity
profile. It is shown in Fig.~\ref{Fig:VelocitySurface} for
$Re=400$ $N=36$ at $x=0$. Such a profile is a consequence of the
braking effect of the magnetic field applied to the electrically
conducting and moving fluid. Shortly, the effect can be understood
in the following way: under the action of the externally applied
magnetic field, electric currents are induced in span-wise
direction. The larger the intensity of the magnetic field ${\bf
B}$, the higher the density of the electric current ${\bf j}$. The
channel walls are insulating and the magnetic field is locally
heterogenous, therefore, to make a closure the electric currents
will leave the area of the high magnetic field (see also the
figures in section \ref{subsec:current}). Then, as ${\bf j}$ and
${\bf B}$ are both present, the Lorentz force ${\bf F}_L = {\bf j}
\times {\bf B}$, hampers stream-wise fluid movement in the bulk of
the channel. The flow  tries to circumvent the area of high
magnetic field as much as possible, and, as a result, the
stream-wise velocity profile will adopt M-shaped  in span-wise
direction, and a stagnant region with stream-wise velocity about
zero (see solid lines in Fig.~\ref{Fig:VelocityProfilex0}) forms
inside the magnetic gap. Various M-shaped  surfaces are given in
[\cite{Sterl:1990}], including their discussion and corresponding
references. It is worth to note that the external magnetic field
selected in [\cite{Sterl:1990}] for the simulation was either
divergence- or curl-free, but not simultaneously divergence- and
curl-free, nevertheless this did not disturb the formation of a
M-shaped  profile.

Fig.~\ref{Fig:VelocityProfilex0} shows the influence of the
interaction parameter $N$ on the M-shaped  velocity profile. The
effect is clear: the higher the value of $N$ the stronger the Lorentz
force braking the flow, the lower the velocity in the center, and thus
the more pronounced the stagnant region.
The solid curves 1 and 2 in Fig.~\ref{Fig:VelocityProfilex0} are
related to the same parameter values $Re=400$, $N=36$, but to
different magnetic fields: the experimental field (curve 1) and the
simplified field (curve 2). This comparison shows that the
character of the flow in this two cases is rather different. In
particular, for the case of simplified field, the width of the
stagnant region is larger but its level is higher than in the case
of the experimental field. This can be explained by the decrease of $B_z$
in span-wise direction for the experimental magnetic field. As the
transverse magnetic field near the side walls consequently for the
experimental field is lower than in the center, the liquid can more easily
flow around the "magnetic obstacle" (see below), therefore the
redistribution of the flow for the experimental magnetic field is more
pronounced than for the simplified field.

\begin{figure}[t]
\begin{center}
    \includegraphics[angle=0, width=3.5in]{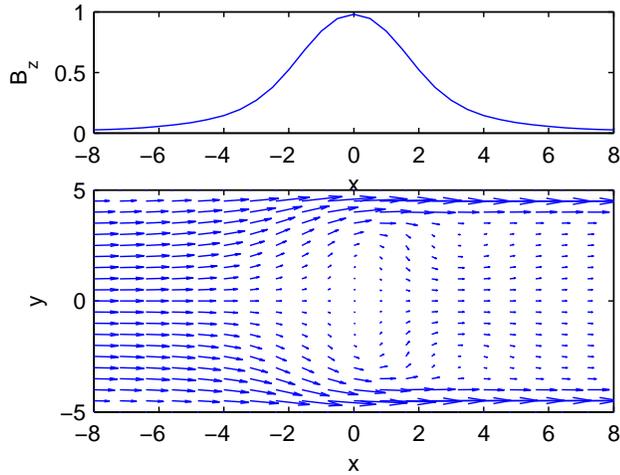} 
    \caption{Vector plot of the velocity field  in the central horizontal
    plane, $z=0$. The upper part shows the intensity of the magnetic field
    on the straight line $z=0$, $y=0$, $Re=400$, $N=36$.}
    \label{Fig:VectorVelocityXY}
\end{center}
\end{figure}

Fig.~\ref{Fig:VectorVelocityXY}.

Another insight for the M-shaped  profile formation might be taken
from the $XY$ vector velocity plot shown in
It was noticed several times a kind of similarity  between the
flow of an electrically conducting liquid passing through an area
of high local heterogenous magnetic field (magnetic obstacle) and
the well known flow around a bluff body (see e.g.
[\cite{Cuevas:Smolentsev:Abdou:2006}]). One can see that the
velocity vectors envelop the central part of the channel nearly in
the same way as it would be for a solid cylindrical obstacle in a
two-dimensional flow. The only difference is that a real body
creates a region with no fluid, say velocity zero, while in the
region with magnetic field there still is fluid, but with very
small velocity.

\begin{figure*}[tb]
\begin{center}
    \includegraphics[angle=0,width=5.0in]{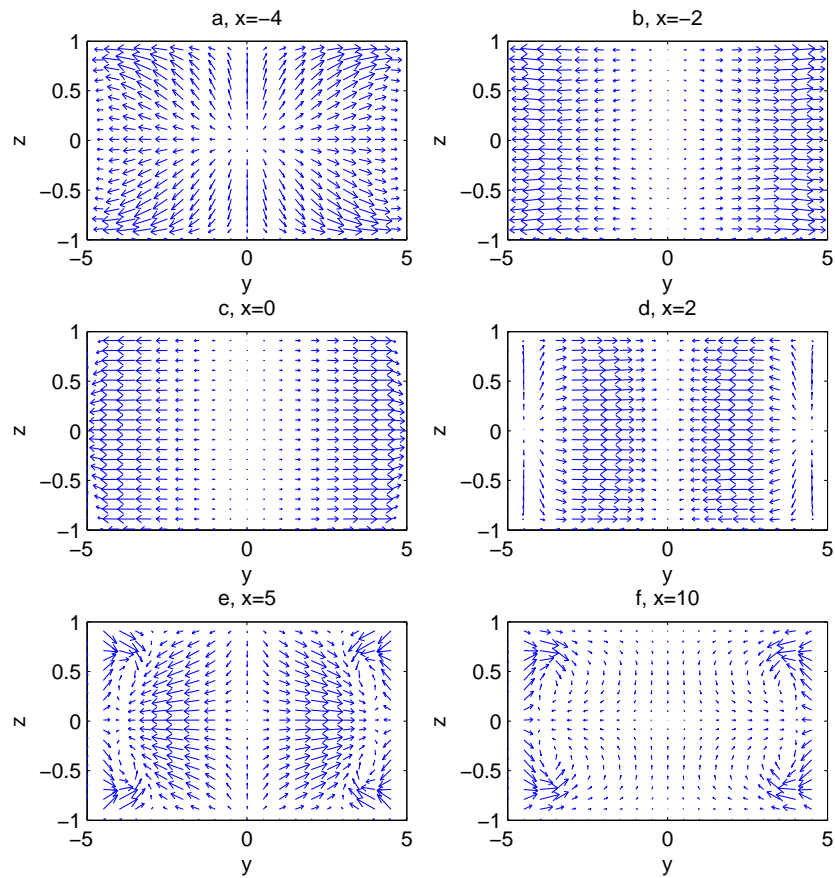} 
    \caption{Vector plots of the vertical and span-wise velocity
    components at the following vertical sections of the channel:
    ($a$) $x=-4$, ($b$) $x=-2$, ($c$) $x=0$,
    ($d$) $x=2$, ($e$) $x=6$, ($f$) $x=11$ at $Re=400$, $N=36$.
    An interesting feature is the swirling flow that arises in the corners
    behind the magnet (plot $e,f$).}
    \label{Fig:VectorVelocityYZ}
\end{center}
\end{figure*}

To analyze span-wise and vertical velocity components we have
plotted a series of vector plots in vertical sections of the
channel at several fixed values of $x$. They are shown in
Fig.~\ref{Fig:VectorVelocityYZ} to highlight the intrinsic
three-dimensional nature of the velocity field. In the region of
increasing magnetic field before the magnet gap (see
Fig.~\ref{Fig:VectorVelocityYZ}$a$), the velocity vector in the
vertical section  is nearly aligned towards the corners of the
channel. This can be explained by the simultaneous action of two
effects: (i) the beginning tendency to form a M-shaped  profile,
and (ii) the formation of Hartmann layers which is accompanied by
a flattening of the velocity profile in vertical direction (see also
Fig.~\ref{Fig:VectorVelocityXZ} below). The second reorganization
process of the flow is accomplished faster, i.e. finished already
at smaller $x$-coordinate, than the first one. This can be
concluded from the second section,
Fig.~\ref{Fig:VectorVelocityYZ}$b$, because the plotted vectors
indicate only a movement in the span-wise direction.

The formation of M-shaped  profile continues further when the flow
passes the maximum of the magnetic field, see
Fig.~\ref{Fig:VectorVelocityYZ}$c$. However, as the flow reaches
the region of decreasing magnetic field, there is an inversion of
the span-wise movement (Fig.~\ref{Fig:VectorVelocityYZ}$d$): now it
is opposite to the one observed before the magnet. These two
processes --- flow towards the side walls at the front of the
magnetic gap and towards the center (i.e., away from the side
walls) after the gap --- demonstrate again that the flow passing a
heterogenous magnetic field has an analogy with the flow around an
obstacle. Increasing further the $x$-value of the vertical section
and thus leaving the region with noticeable magnetic field, the
Hartmann profile begins to transform back into a parabolic
profile. Consequently, there must be a {\em vertical} movement from the
bottom and top walls towards the center, see
Fig.~\ref{Fig:VectorVelocityYZ}$e$ (note: we are only speaking about
the vertical velocity component; the span-wise velocity
component in section $e$ is directed away from the center, compare
with Fig.~\ref{Fig:VectorVelocityXY}).
This phenomenon is opposite
to that observed in Fig.~\ref{Fig:VectorVelocityYZ}$a$: now there
is no more braking force keeping the velocity profile flattened in
the vertical direction. As a result, the flow develops a swirling
component in the channel corners. The swirling behavior extends
far behind the magnet, see Fig.~\ref{Fig:VectorVelocityYZ}$f$. To
our knowledge, this effect of swirling flow in the corners inside
the rectangular channel after passing the region of heterogeneous
magnetic field has not been mentioned before elsewhere.


\begin{figure}[tb]
\begin{center}
    \includegraphics[angle=0,width=3.6in]{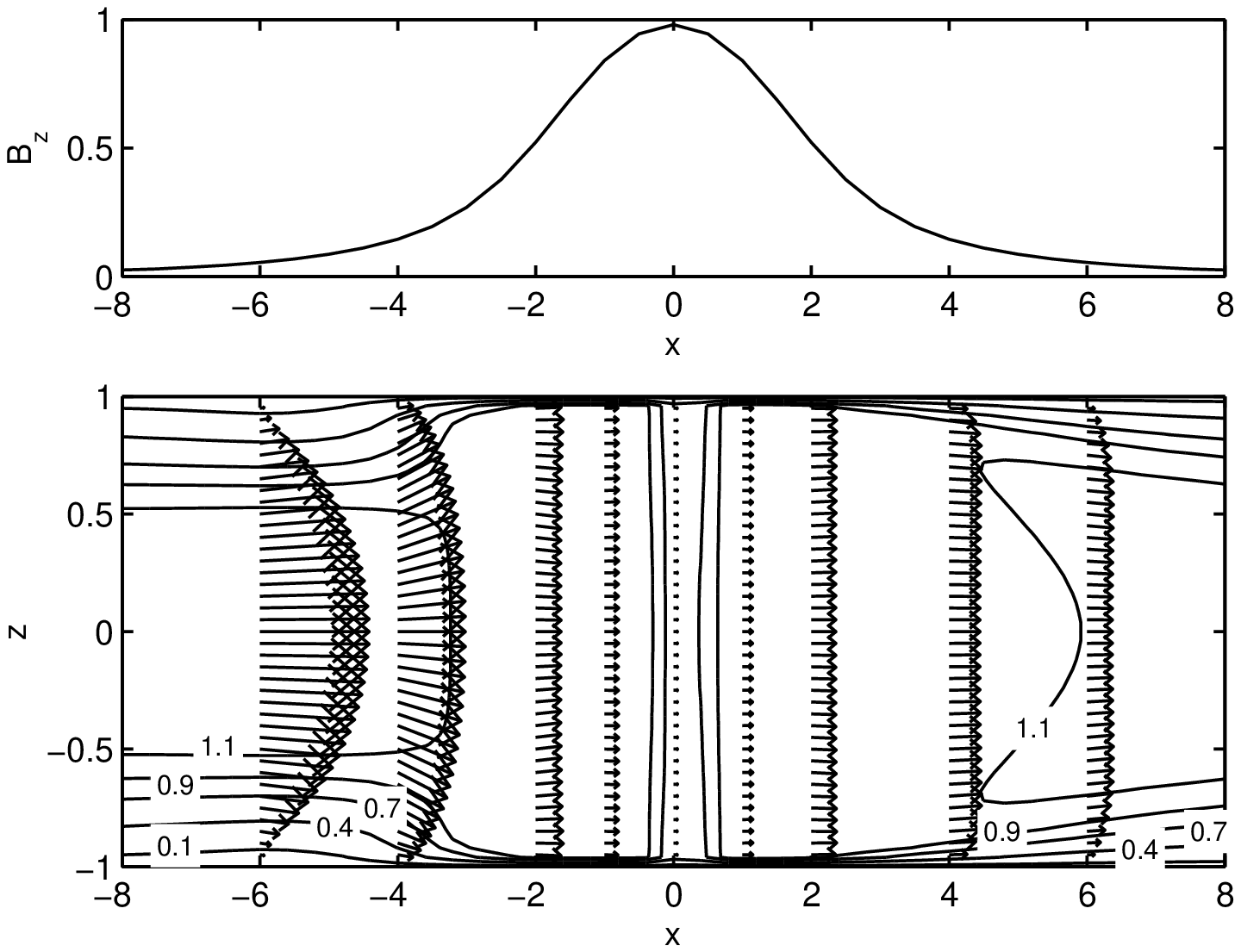} 
    \caption{Vector plot of the velocity components $u_x$ and $u_z$ in the
    central vertical plane at $y=0$, $Re=400$, $N=36$.
    The isolines are plotted at constant level of the normalized vertical
    profile $p(z;x)$ and give additional information about the
    flatness of the velocity profile.}
    \label{Fig:VectorVelocityXZ}
\end{center}
\end{figure}

To get more insight how the vertical velocity redistribution takes
place, Fig.~\ref{Fig:VectorVelocityXZ} presents a vertical section along
the midplane in stream-wise direction of the channel. The velocity
vectors shown are the projection $(u_x,u_z)$ of the total velocity.
One clearly sees how the vertical velocity profile flattens before
the magnet and becomes more convex after the magnet.
Interestingly, the flattening starts in the center of the profile
because of the action of the Lorentz force while the decay of the
Hartmann layers begins at the top and the bottom wall because of
the wall friction and viscosity in the fluid.
The isolines of the normalized vertical profile $p(z;x)$ in
Fig.~\ref{Fig:VectorVelocityXZ} give an additional information to
judge how flat the profile is at a given stream-wise coordinate $x$.
We define the normalized vertical profile as
$p(z;x)=u_x(x,0,z)/u_0(x)$, where $u_0(x)=\frac{1}{2}\int_{-1}^{1}
u_x(x,0,z)dz$ is the mean velocity in the vertical middle plane at given
$x$ and $y=0$.
Horizontal isolines in Fig.~\ref{Fig:VectorVelocityXZ} correspond to a
profile, where the velocity changes in the vertical direction as for
the parabolic profile in the inflow region, while the almost vertical
isolines under the magnet represent regions with a flat profile, i.e.
the Hartmann profile.
Before (behind) the magnet the isolines converge (diverge)
which demonstrates again the flattening (de-flattening) of the
vertical velocity profile. The flattening process turns out to
take place much faster (in four length units) than the de-flattening
which is far from being finished in the six shown length units
after the magnetic gap.

\subsection{\label{subsec:epot}Electric potential}

\begin{figure}[t]
\begin{center}
    \includegraphics[angle=0, width=3.2in]{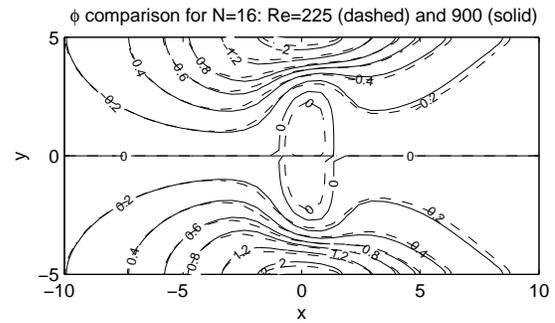} 
    \caption{Electric potential distribution, $\phi(x,y)$ in central
    horizontal plane ($z=0$) for $N=16$, simulation: $Re=900$ (solid)
    and $R=225$ (dashed).}
    \label{Fig:PotentialContourN16}
\end{center}
\end{figure}

One main focus of this work is the demonstration that the
flow under the magnet is determined by the value of the interaction
parameter $N$, independently of the individual values of the Hartmann and
Reynolds numbers as long as the Reynolds number is high enough.
For this aim we first start
with a comparison of two numerical computations for the same interaction
number $N$ but with different Reynolds numbers.
Fig.~\ref{Fig:PotentialContourN16} shows such a comparison
at $N=16$ for $Re=900$ (solid lines) and $Re=225$ (dashed lines).
One can actually see that the two contour line sets are not prominently
distinct despite the fact that for both cases the $Re$ numbers differ
by a factor of 4, and the $Ha$ numbers differ by a factor of 2. Before the
magnetic gap the dashed and solid lines closely coincide with each
other, and after the magnets there is a slight discrepancy since
the inertial force for $Re=900$ is higher than for $Re=225$. More
discrepancies one can find in the thicknesses of viscous boundary
layers, however they are essentially of no importance since we are
interested in the processes taking place in the central part of
the magnetic gap.

Now let us explain qualitatively the electric field distribution
in the system under consideration. In the quasi-static approximation
the current according to Ohm's law eq.~(\ref{eq:NSE:Ohm}) is the sum
of two terms: the electric field induced by the motion of the conducting
fluid volume inside the magnetic field, ${\bf u}\times{\bf B}$, and
the electric field ${\bf E}=-\nabla\phi$ derived from an electric
potential $\phi$ generated inside the fluid volume because of
the solenoidality of the electric current (due to Amp\`ere's
law $\nabla \times {\bf B}=\mu {\bf j}$) and the isolating boundary
conditions at the walls forcing currents to close inside the fluid
volume. Taking the divergence of (\ref{eq:NSE:Ohm}) one gets the
Poisson equation (\ref{eq:NSE:Poisson}) for which the right hand side
$\nabla\cdot({\bf u}\times{\bf B}) = {\bf B}\cdot{\bf w}$ plays the
role of the inhomogeneity\footnote{The second term ${\bf u} \cdot
(\nabla \times {\bf B})$ vanishes because the fluid volume is outside
of the external magnet and induced magnetic fields are neglected
in the quasi-static approximation.}.
Comparing with the Poisson equation for usual electrodynamics,
$\triangle\phi({\bf r})=-\rho({\bf r})/(\varepsilon_0\varepsilon)$,
one sees that the expression
$\rho^*=-\varepsilon_0\varepsilon \, {\bf B}\cdot{\bf w}$
can be considered formally as an electric charge density induced by the
movement of the electrically conducting fluid in the magnetic field.
In the duct flow, the largest contributions to ${\bf B}\cdot{\bf w}$
are generated with the predominant vertical component $B_z$ of the
magnetic field together with the span-wise velocity gradient
$\partial u_x/\partial y$ of the stream-wise velocity component, i.e.
$\rho^*\approx \varepsilon_0\varepsilon B_z (\partial u_x/\partial y)$.
Therefore, looking in stream-wise direction, one finds a negative charge
density near the left wall and a positive charge density near the right
wall of the channel which correspond to the outer flanks of the
M shape profile under the magnet. The electric field created by this
charge density inside of the channel points in span-wise direction
\textit{parallel} to the  $y$ axis\footnote{Two remarks: (1) The
electric field direction coincides, of course, with the direction of
${\bf u}\times {\bf B}$.
(2) Directly at the walls the charge density is high, but the induced
electric field has to be zero because of the isolating boundary
condition.}. (The $y$ axis points from the right to the left wall of
the channel).
The braking action of the Lorentz force leads to a strongly
diminished flow in the central region under the magnet and in this
way the M shape profile is created. The inner flanks of the M shape
profile provide opposite velocity gradients and opposite charge
densities causing an electric field \textit{antiparallel} to the
$y$-axis in the central region.
It depends now on the strength of the magnetic field and
the corresponding braking effect, whether there exists a region in
the middle of the channel under the magnet, where the total electric
field ---  the sum of both described charge densities from the outer and
inner flank of the M shape profile --- is antiparallel. In this
case the electric field has to change its sign two times along the
span-wise direction, corresponding to three sign changes for the electric
potential. In the contour
plot of the electric potential, the stagnant region is manifested
by closed contour lines as we shall see below in
Fig.~\ref{Fig:PotentialContourN20},~\ref{Fig:PotentialContourN10}.

\begin{figure}[t]
\begin{center}
    \includegraphics[angle=0, width=3.3in]{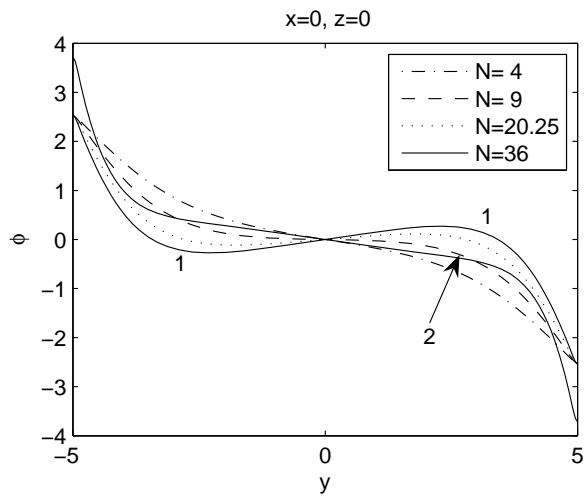} 
    \caption{Simulated span-wise electric potential profiles in the center of
    the magnetic  gap ($x=0$ and $z=0$), $Re=400$ and  $N=4$ (dot-dashed),
    $N=9$ (dashed), $N=20.25$ (dotted),  $N=36$, experimental magnetic field (solid line 1)
    and  $N=36$, simplified field (solid line 2).}
    \label{Fig:PotentialProfilex0}
\end{center}
\end{figure}

To illustrate the previous deduction about electric potential
behavior we plotted in Fig.~\ref{Fig:PotentialProfilex0} span-wise
profiles of $\phi$ for different interaction parameter $N$. One
can see that for the low interaction parameter $N=4$ (dot-dashed)
$\phi(y)$ shows clearly monotonic behavior, while for the high
interaction parameter $N=36$ (solid line 1) the sign of the electric
potential changes three times. Corresponding curves
in Fig.~\ref{Fig:VelocityProfilex0} and
\ref{Fig:PotentialProfilex0} are plotted with the same line types.
One clearly observes the tendency: the development of the
inflection of the $\phi(y)$ curves
(Fig.~\ref{Fig:PotentialProfilex0}) is accompanied by a lower
minimum of the velocity profile (Fig.~\ref{Fig:VelocityProfilex0})
around $y=0$.

\begin{figure*}[t]
\begin{center}
  \subfigure[]{\label{Fig:PotentialProfileN20}
    \includegraphics[angle=0, width=2.9in]{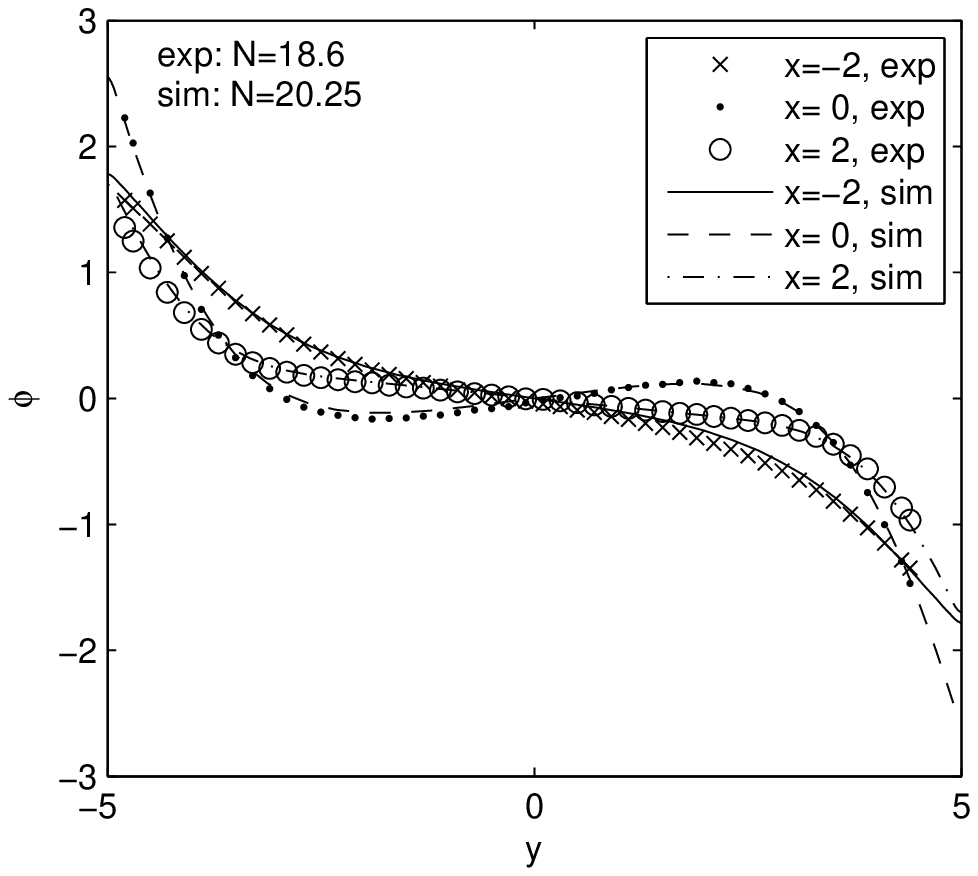}}
  \subfigure[]{\label{Fig:PotentialProfileN10}
    \includegraphics[angle=0, width=3.0in]{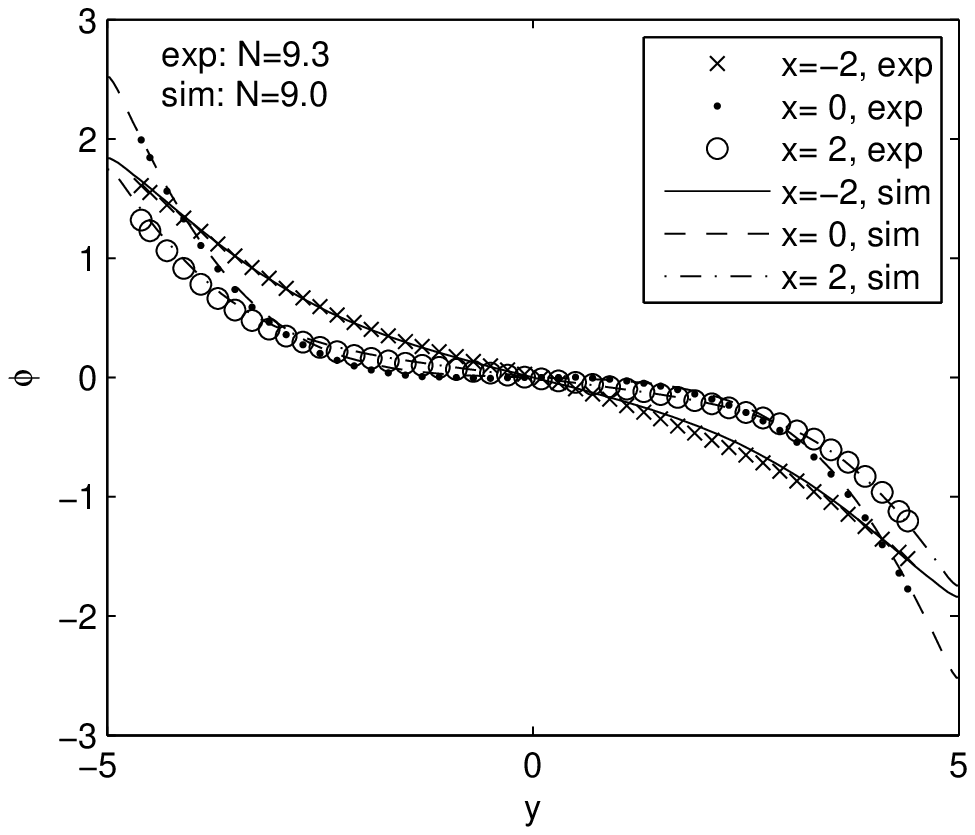}}
  \caption{Experimental (symbols) and simulated
    (lines) span-wise electric potential profiles at $z=0$ and
    $x=-2$ (crosses and solid lines), 0 (dots and dashed lines), 2 (circles and dot-dashed lines).
    Plot $a$ is for $N\approx 20$: $(Re,N)=(2000,18.6)$ (experiment) and $(Re,N)=(400,20.25)$ (simulation).
    Plot $b$ is for $N\approx 9$: $(Re,N)=(4000,9.3)$ (experiment) and $(Re,N)=(400,9)$ (simulation).}
\end{center}
\end{figure*}

The next two figures show how the electric potential changes by
passing the magnetic field region. They are  given in comparison
with the experimental results. Fig.~\ref{Fig:PotentialProfileN20}
shows the profiles for $N\approx 20$ and
Fig.~\ref{Fig:PotentialProfileN10} shows the profiles for
$N\approx 9$.  As it should be, kinks of the electric potential in
Fig.\ref{Fig:PotentialProfileN20} are more expressive than in
Fig.\ref{Fig:PotentialProfileN10} due to the larger interaction
parameter. Also, the figures demonstrate that this effect is most
expressive in the center of the magnetic gap ($x=0$) where the
magnetic field and, therefore, the braking Lorentz force is
maximal. In the region of the inward ($x=-2$) and outward ($x=2$)
magnetic field gradient, electric potential behavior is monotonic.
The comparison between the simulated and experimental results
reveals almost perfect accordance.

%

To complete the comparison with the experimental
[\cite{Andrejew:Kolesnikov:2006}] data we present also the contour
plots. Fig.~\ref{Fig:PotentialContourN20} shows the experimental
and numerical results of the electric potential distribution at
$N\approx 20$ . Two contour plots give the level lines of the
electric potential, the first, as it was directly and
systematically measured in the experiment, and the second, as
computed numerically. Qualitatively, the electric potential
distribution has the following main features: (i) it is an
antisymmetric Al contour map with respect to the axis $y=0$ (ii)
there are two global extrema directly at the side walls exactly
under the peak of the magnetic field ($x=0$) and (iii) there are
two local sign alternating extrema slightly shifted in stream-wise
direction at one third distance from the side wall (measured by
the total span-wise width of the channel),
Fig.~\ref{Fig:PotentialContourN20}. As one can see, all the
features of the experimental electric potential measured at
$Re=2000$ are excellently  reflected in the simulation at $Re=400$
since both have the similar interaction parameter, $N\approx 20$.
Another comparison for $N\approx 9$ is given in
Fig.~\ref{Fig:PotentialContourN10}. Now, the closed lines of the
electric potential are weaker because of the lower interaction
parameter. The good accordance between experimental and simulated
results is observed in Fig.~\ref{Fig:PotentialContourN10} as
well.

\begin{figure}[t]
\begin{center}
    \includegraphics[angle=0, width=3.3in]{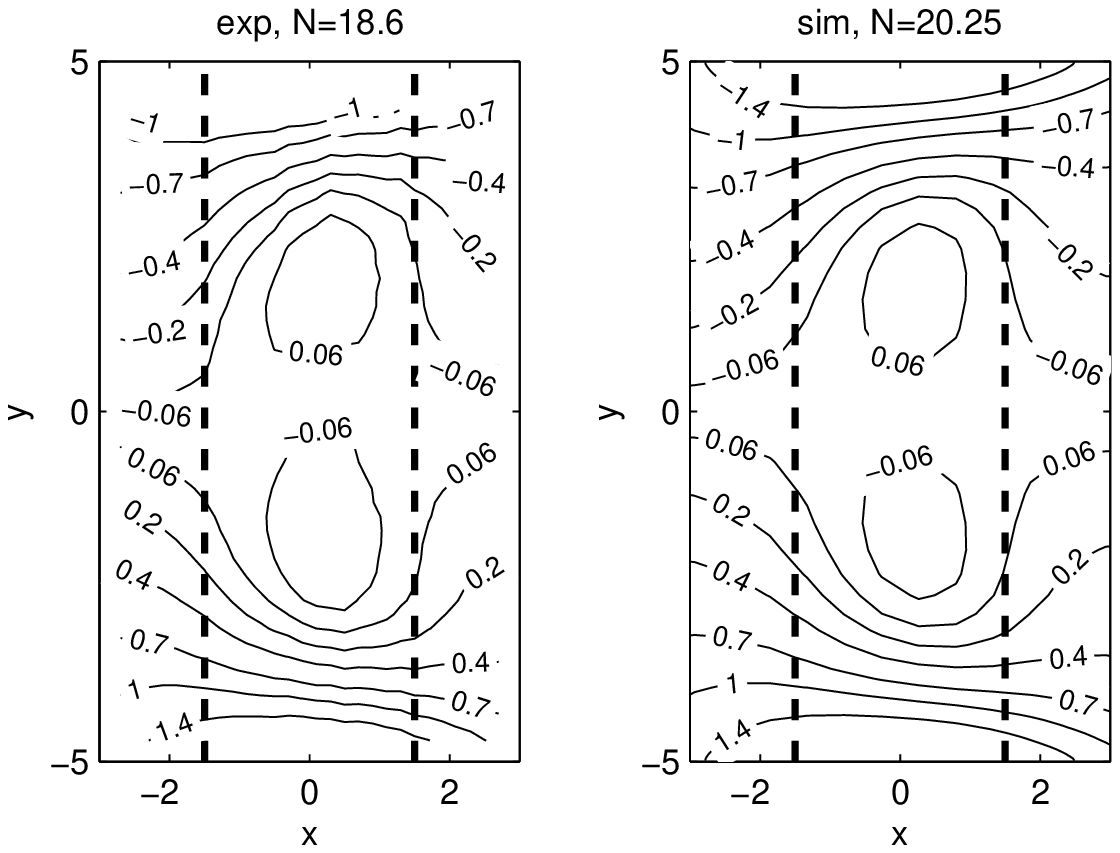} 
    \caption{Electric potential distribution, $\phi(x,y)$, in the central
    horizontal plane ($z=0$) for $N\approx 20$: experiment, $Re=2000$, $N=18.6$ (left)
    and simulation $Re=400$, $N=20.25$ (right). Bold dashed lines are physical
    borders of the magnetic gap.}
    \label{Fig:PotentialContourN20}
\end{center}
\end{figure}

\begin{figure}[t]
\begin{center}
    \includegraphics[angle=0, width=3.3in]{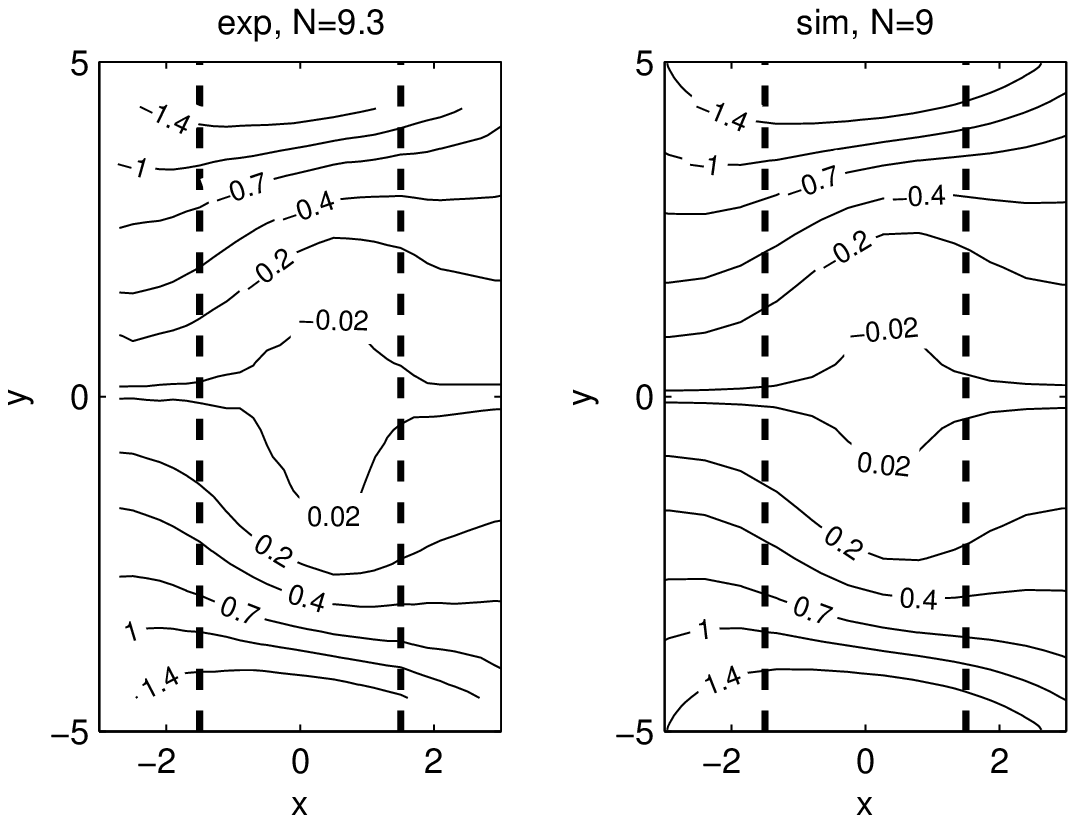} 
    \caption{Electric potential distribution, $\phi(x,y)$ in central
    horizontal plane ($z=0$) for $N\approx 9$: experiment, $Re=4000, N=9.3$ (left)
    and simulation $Re=400$, $N=9$ (right). Bold dashed lines are physical
    borders of the magnetic gap.}
    \label{Fig:PotentialContourN10}
\end{center}
\end{figure}

Summing up, we find a very good agreement between the
experimental and the numerical results comparing data for
the same interaction parameter $N$. This holds even in complementary
regimes with respect to the Reynolds number: turbulent inflow in the
experiment, and laminar calculations in the simulation. This shows that
the interaction parameter indeed
is the governing parameter for the flow under and near to the
magnets, and that the flow in this region is determined by the magnetic
and inertia forces. If the Reynolds number is not high enough, the
viscous force begins to play a role as a third force in the system
as can be observed to a small extent in our first comparison, see
Fig.~\ref{Fig:PotentialContourN16}, of this subsection.



\begin{figure}[t]
\begin{center}
    \includegraphics[angle=0, width=3.3in]{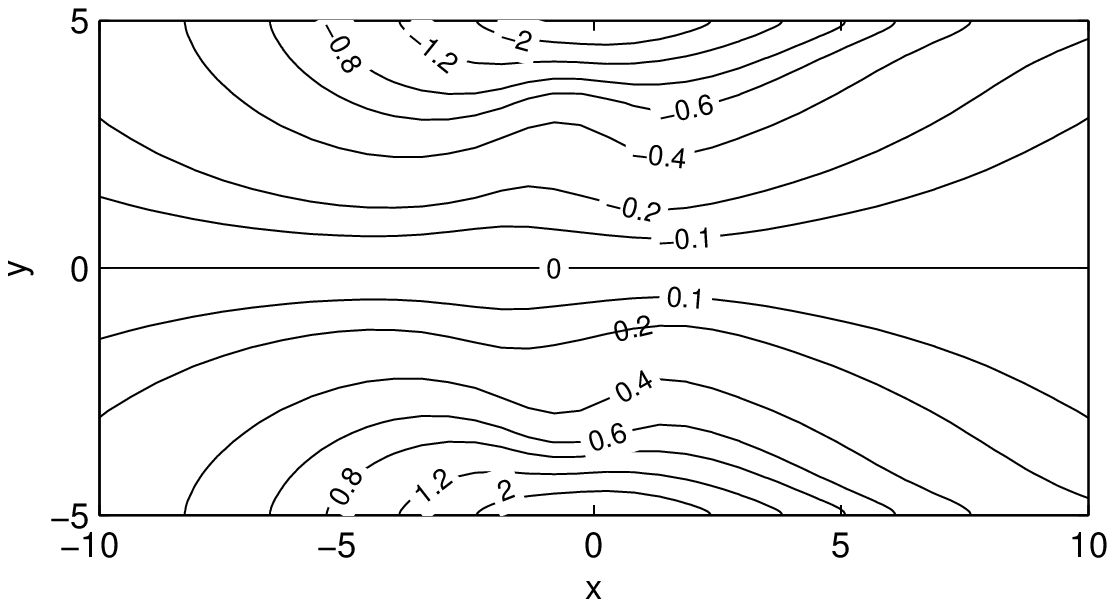} 
    \caption{Electric potential distribution, $\phi(x,y)$ in central
    horizontal plane $z=0$ for $Re=400$, $N=36$, and magnetic field without
    span-wise dependence. }
    \label{Fig:PotentialContourNoY}
\end{center}
\end{figure}

\egbert{replacement of the part, which I found very
incomprehensible. I think, I got what was intended to be said.} As
the electric potential does not show a visible reaction on
turbulence or no turbulence in the inflow, one could conclude that
it is a  quantity which is rather insensitive to different
influences. We will show now that this is not the case, and
consider for this aim the action of the span-wise decrease of the
experimental magnetic field on the electric potential distribution.
\egbert{end replacement.} This dependence is especially
interesting, since many people believe that in most cases only the
transverse component of the magnetic field, changing along
stream-wise coordinate, is of importance. For instance, most of the
numerical simulations were performed only with $B_z(x)$
dependence, see, e.g. [\cite{Sterl:1990, Alboussiere:2004}]. To
clarify the role of the magnetic field and demonstrate the
sensitivity of the $\phi(x,y)$ map, we fixed $(Re,N)=(400,36)$ and
compare two variants  with the real experimental (showing
$y$-dependence) and simplified (span-wise uniform) magnetic field.
As it is supposed typically, the simplified magnetic field has
only stream-wise dependence in its intensity, and is constant at
every fixed $x$ in span-wise direction. The simulated results for
the simplified field are presented in
Fig.~\ref{Fig:PotentialContourNoY}. One discovers that now the
picture changes even qualitatively: there are no more closed lines
of the electric potential, that is, the potential drops
monotonically from one side wall to another (compare also the
solid curves 1 and 2 in Fig.~\ref{Fig:PotentialProfilex0}). This
is a strong indication that the factors, which are decisive for
the comparison of experimental and simulated results, are the
maximal interaction parameter in the magnetic gap and the proper
configuration of the magnetic field.

\subsection{\label{subsec:current}Electric currents paths}

\begin{figure}[t]
\begin{center}
    \includegraphics[angle=0, width=3.6in]{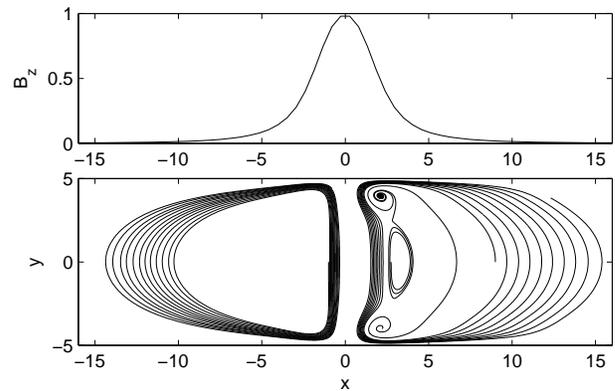}
    \caption{Electric current paths in the central horizontal
       plane ($z=0$) under the experimental magnetic field, $Re=400$,
       $N=36$. The lines are not equidistant since they were obtained
       by integration from a few starting points at $(x,y)=(-0.5,0),
       (0.5,0), (2.1, 4.1), (2.1, -4.1)$.}
    \label{Fig:CurrentContourXY}
\end{center}
\end{figure}

Fig.~\ref{Fig:CurrentContourXY} shows the closure of the electric
currents in the central horizontal plane ($z=0$) which were
calculated as lines tangent at every point to the vector of
electric current $\bf{j}$. For sake of simplicity only few paths
are shown, one in the region of increasing magnetic field, and
others after the magnet in the region of decreasing field. In the
case of constant magnetic field the loops of the electric current
are located entirely in vertical planes and most of the current is
concentrated in the Hartmann layers. By contrast, in the case of
the heterogenous magnetic field, since an electric current intends
to close itself in the region where magnetic field is minimal
(in order to close along paths with smallest resistance), one
sees the current loops close themselves in the horizontal planes.
The characteristic length of the loops is rather large as can be
seen in Fig.~\ref{Fig:CurrentContourXY}: the turning point of
$\bf{j}$, starting under the peak of the magnetic field ($x\approx
0$), is at $|x|\approx 10...15$, while a remarkable magnetic field
intensity is felt only up to $x\approx 5$. This fact must be taken
into consideration when one selects where to put inlet and outlet
of the numerical simulation box, otherwise it is possible to get
artificial findings.

\begin{figure}[t]
\begin{center}
    \includegraphics[angle=0, width=3.5in]{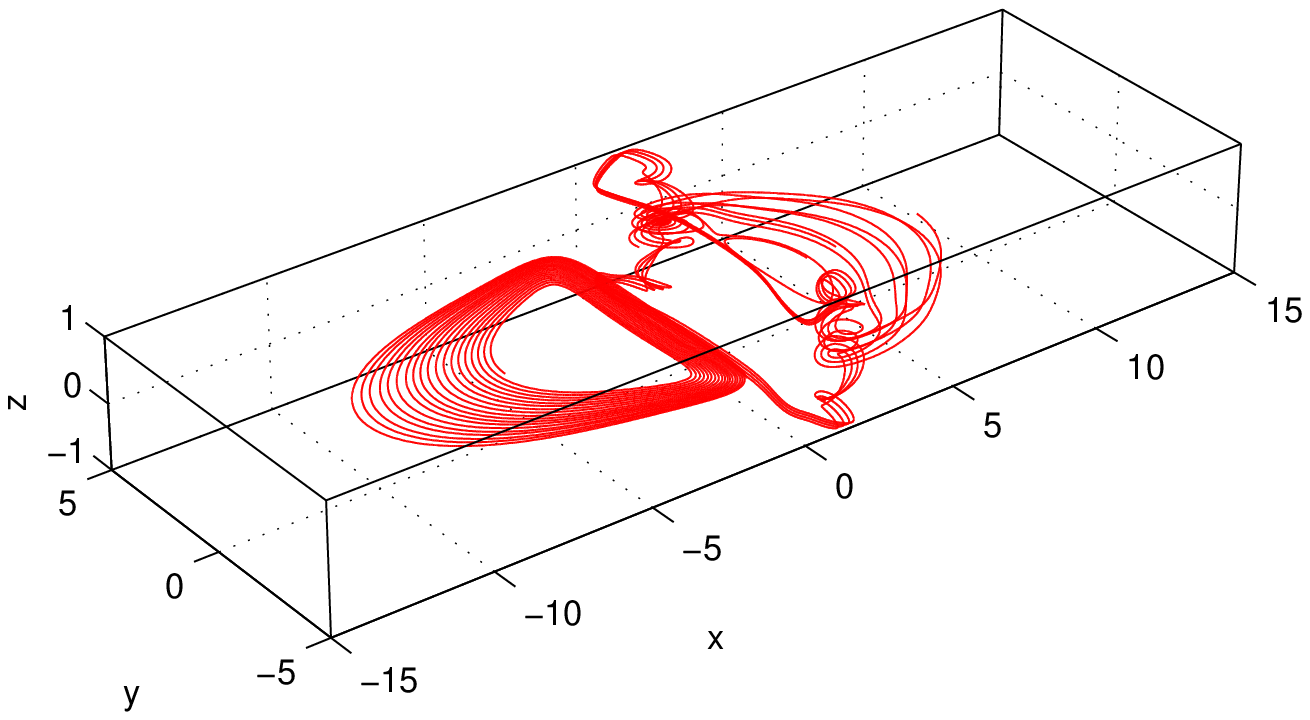}
    \caption{3D electric current path for the experimental magnetic field.}
    \label{Fig:Currents3Dbrick5}
\end{center}
\end{figure}

Fig.~\ref{Fig:CurrentContourXY} reveals small closed loops for the
electric current at $x\approx 2.2$ and $y\approx \pm 4.1$. These
loops are interesting because of the closure of the electric current
is typically a big horizontal loop which envelops the space of the
magnetic field gradient. The geometric explanation for these small exotic
loops is that they are projections to the $XY$ plane from complicated
electric paths developing essentially in 3D space.
Fig.~\ref{Fig:Currents3Dbrick5} clearly shows these 3D paths:
they are helices connecting the Hartmann layer and the middle
horizontal plane.

\begin{figure}[t]
\begin{center}
    \includegraphics[angle=0, width=3.5in]{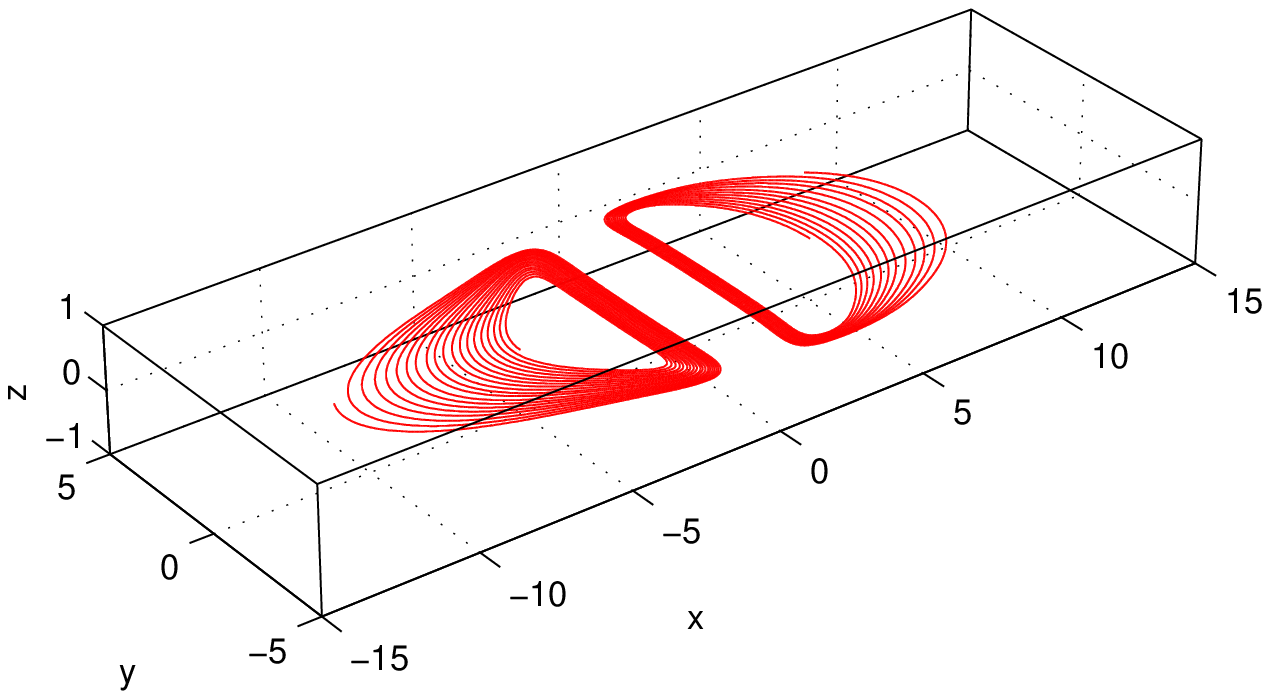}
    \caption{3D electric current path for the simplified  magnetic
    field without span-wise dependence.}
    \label{Fig:Currents3Dbrick99}
\end{center}
\end{figure}


As well in our calculation presented in this paper, see
Fig.~\ref{Fig:Currents3Dbrick5}, as also in
other calculations which are not shown here, helical current paths
are only present when the level lines of the electric potential relief
in the horizontal midplane contains closed lines, see
Fig.~\ref{Fig:PotentialContourN20}. Our results strongly suggest that
the presence of closed lines of the electric potential is a necessary
condition for the appearance of helical current paths. As the
span-wise uniform magnetic field does not lead to a triple change of
the span-wise electric potential profile (i.e. no closed lines in
Fig.~\ref{Fig:PotentialContourNoY}), the 3D paths of the electric
currents for this case consequently are simple closures in the
horizontal plane as is shown in Fig.~\ref{Fig:Currents3Dbrick99}.

There are also correlations between the span-wise inhomogeneity of
the magnetic field, 3D helical currents, and features of the
stagnant region in M-shaped  velocity profiles. One sees a broad
constant plateau in $u_x(y)$ for the simplified field while for
the experimental field there is a degraded minimum approaching zero
(compare solid curves 1 and 2 in
Fig.~\ref{Fig:VelocityProfilex0}). Since the level of the stagnant
plateau for the span-wise uniform magnetic field is sufficiently
high to keep electromotive contribution ${\bf u} \times {\bf B}$
in the electric current ${\bf j}$, there is no double sign change
of the electric potential $\phi(y)$ (compare solid curves 1 and 2
in Fig.~\ref{Fig:PotentialProfilex0}). The stagnant plateau is not
so broad but considerably lower for the experimental magnetic field,
because of the lower magnetic field strength near the side walls.
The latter induces the braking Lorentz force near side walls to be
smaller than in the center. As a result, the case of the real
field looks somehow as a flow around the magnetic obstacle, while
the simplified field is looks more similar to an uniform
semi-penetrable barrier.


The 3D helices of the electric current in the case of the real
field (Fig.~\ref{Fig:Currents3Dbrick5}) arise near the turning
points where the electric field $\bf E$ changes its sign, hence
$E \sim 0$. These helices are maintained mainly by the
electromotive component (${\bf u}\times {\bf B}$) in Ohm's law,
eq.~\ref{eq:NSE:Ohm}. On the other hand, these helices are located
in the region of space where stream-wise component $B_x(x)$ reaches
the largest value, see Fig.~\ref{Fig:ProfilesMagneticField}, in
the center of the outward magnetic field region. Referring to the
vector product ${\bf u}\times{\bf B}$, we see that the term $u_y
B_x$ is of importance for the size of the $j_z$ component, which
is responsible for the uprise of the helix. Thus, in order to
catch the helices one has to keep in the simulation all the
components of the magnetic field.

\section{Conclusions}

We have carried out numerical simulations for liquid metal channel
flow under inhomogeneous magnetic field. For the computations the
same channel geometry and the magnetic field configuration as in
the experiment of [\cite{Andrejew:Kolesnikov:2006}] were used.

Computations of the velocity field showed that all known general
features --- like for example the formation of Hartmann layers at
the walls perpendicular to the main component of the magnetic
field, or the formation of an M shaped profile in span-wise
direction --- are represented correctly by our numerical code. All
features of the flow together lead to a complex three-dimensional
flow structure, which for the investigated regime of interaction
parameters, $4\leq N \leq 36$, can only be determined numerically.
As a new feature of the velocity field a swirling flow in the corners
of the duct is observed. It begins shortly after the magnetic
gap and extends far into the outflow region.

The main goal of this work was to compare the electric potential
distribution measured in the experiment with that of our numerical
simulation. We were able to find a very good agreement of the
electric potential distribution for two sets of parameters which
were used in the experiment ((i)~$N=18.6$, $Re=2000$,  and
(ii)~$N=9.3$ $Re=4000$) and the corresponding sets ((i)~$N=20.25$,
$Re=400$, and (ii)~$N=9$, $Re=400$) in our numerical calculations.
This comparison shows that the electric potential distribution in
the magnetic field region is solely determined by the value of the
interaction parameter. This makes sense as in the magnetic field
region the Lorentz force and the inertial force are strongest,
while the viscous force is only important in the Hartmann layers
and near to the side walls. The turbulence, which is present in
the inflow of the experimental channel is in the magnetic field
region already negligible and has no influence on the electric
potential distribution. A numerical test with a different inflow
profile, which is more flat like a turbulent profile, also showed
no difference.

In addition, we could demonstrate that local extrema of the
electric potential appear for interaction parameter higher than a
critical value, which lies between 9 and 18. Simulations with a
simplified magnetic field (only dependent on vertical and
stream-wise coordinate, no span-wise field component) showed that
local extrema of the electric potential map totally vanish in this
case. This means that already a slight variation of the
inhomogeneous magnetic field can have a strong influence of the
electric potential distribution. The same sensitivity on the
magnetic field structure is found for the electric current
density: In the case of the experimental magnetic field our computations
revealed the appearance of complicated helical current lines which
are not present for the simplified magnetic field.

\acknowledge{ The authors express their gratitude to the Deutsche
Forschungsgemeinschaft for financial support in the frame of the
"Research Group Magnetofluiddynamics" at the Ilmenau University of
Technology under grant ZI 667/2-3. The simulations were carried
out on a JUMP supercomputer, access to which was provided by the
John von Neumann Institute (NIC) at the Forschungszentrum
J\"{u}lich. We are grateful for many fruitful discussions with Andre
Thess. Special thanks go to our experimental colleges, Yuri
Kolesnikov and Oleg Andreev, for an always close exchange of thoughts and
providing the experimental data to compare with our numerical
results. As well, the authors appreciate the scientific exchange
on our subject with Yves Delannoy and Jacqueline Etay in Grenoble,
France, which was supported by the Deutscher Akademischer Austauschdienst. }

\bibliography{Channel_FDMP}

\begin{thebibliography}{19}
\expandafter\ifx\csname natexlab\endcsname\relax\def\natexlab#1{#1}\fi

\bibitem[Alboussiere(2004)]{Alboussiere:2004}
\textbf{Alboussiere, T.} (2004): \ignore.
\newblock A geostrophic-like model for large {H}artmann number flows.
\newblock \emph{ J. Fluid Mech}, vol. 521, pp. 125--154.

\bibitem[Andreev, Kolesnikov, and Thess(2006)Andreev, Kolesnikov, and
  Thess]{Andrejew:Kolesnikov:2006}
\textbf{Andreev, O.; Kolesnikov, Y.; Thess, A.} (2006): \ignore.
\newblock Experimental study of liquid metal channel flow under the influence
  of a non-uniform magnetic field.
\newblock \emph{ Phys. Fluids}, vol. 18, pp. 065108.

\bibitem[Cuevas, Smolentsev, and Abdou(2006)Cuevas, Smolentsev, and
  Abdou]{Cuevas:Smolentsev:Abdou:2006}
\textbf{Cuevas, S.; Smolentsev, S.; Abdou, M.} (2006): \ignore.
\newblock On the flow past a magnetic obstacle.
\newblock \emph{ JFM}, vol. 553, pp. 227 -- 252.

\bibitem[Davidson(1999)]{Davidson:Review:1999}
\textbf{Davidson, P.} (1999): \ignore.
\newblock Magnetohydrodynamics in {M}aterials {P}rocessing.
\newblock \emph{ Annual Review of Fluid Mechanics}, vol. 31, pp. 273--300.

\bibitem[Griebel, Dornseifer, and Neunhoeffer(1995)Griebel, Dornseifer, and
  Neunhoeffer]{Griebel:book:1995}
\textbf{Griebel, M.; Dornseifer, T.; Neunhoeffer, T.} (1995): \ignore.
\newblock \emph{ Numerische Str{\"o}mungssimulation in der
  Str{\"o}mungsmechanik}.
\newblock Vieweg Verlag, Braunschweig.

\bibitem[Hirsch(1988)]{Hirsch:1988}
\textbf{Hirsch, C.} (1988): \ignore.
\newblock \emph{ Numerical computation of internal and external flows. Volume I
  and II}.
\newblock John Wiley and Sons, Chichester.

\bibitem[Kenjerec and Hanjalic(2000)]{Kenjerec:Hanjalic:2000}
\textbf{Kenjerec, S.; Hanjalic, A.} (2000): \ignore.
\newblock On the implementation of effects of {L}orentz force in turbulence
  closure models.
\newblock \emph{ Int. J. of Heat and Fluid Flow}, vol. 21, pp. 329--337.

\bibitem[Kenjerec and Hanjalic(2004)]{Kenjerec:Hanjalic:2004}
\textbf{Kenjerec, S.; Hanjalic, A.} (2004): \ignore.
\newblock A direct-numerical-simulation-based second-moment closure for
  turbulent magnetohydrodynamic flows.
\newblock \emph{ Physics of Fluids}, vol. 16, no. 5, pp. 1229--1241.

\bibitem[Knaepen and Moin(2004)]{Knaepen:Moin:2004}
\textbf{Knaepen, B.; Moin, P.} (2004): \ignore.
\newblock Large-eddy simulation of conductive flows at low magnetic {R}eynolds
  number.
\newblock \emph{ Physics of Fluids}, vol. 16, no. 5, pp. 1255--1261.

\bibitem[Lavrentiev, Molokov, Sidorenkov, and Shishko(1990)Lavrentiev, Molokov,
  Sidorenkov, and Shishko]{Lavrentiev:Molokov:Magnetohydrodynamics:1990}
\textbf{Lavrentiev, I.; Molokov, S.; Sidorenkov, S.; Shishko, A.} (1990):
  \ignore.
\newblock Stokes flow in a rectangular magnetohydrodynamic channel with
  nonconducting walls within a nonuniform magnetic field at large {H}artmann
  numbers.
\newblock \emph{ Magnetohydrodynamics}, vol. 26, no. 3, pp. 328--338.

\bibitem[Peyret and Taylor(1983)]{Peyret:Taylor:1983}
\textbf{Peyret, R.; Taylor, T.} (1983): \ignore.
\newblock \emph{ Computational methods for fluid flow}.
\newblock Springer--Verlag, New York.

\bibitem[Roberts(1967)]{Roberts:1967}
\textbf{Roberts, P.~H.} (1967): \ignore.
\newblock \emph{ An introduction to {M}agnetohydrodynamics}.
\newblock Longmans, Green, New {Y}ork.

\bibitem[Sellers and Walker(1999)]{Sellers:Walker:1999}
\textbf{Sellers, C.; Walker, J.} (1999): \ignore.
\newblock Liquid-metal flow in an electrically insulated rectangular duct with
  a non-uniform magnetic field.
\newblock \emph{ Int. J. Eng. Sci.}, vol. 37, pp. 541--552.

\bibitem[Sommeria and Moreau(1982)]{Sommeria:Moreau:MHDturblulence:1982}
\textbf{Sommeria, J.; Moreau, R.} (1982): \ignore.
\newblock Why, how, and when, {MHD} turbulence becomes two-dimensional.
\newblock \emph{ J. Fluid Mech}, vol. 118, pp. 507--518.

\bibitem[Sterl(1990)]{Sterl:1990}
\textbf{Sterl, A.} (1990): \ignore.
\newblock Numerical simulation of liquid-metal {MHD} flows in rectangular
  ducts.
\newblock \emph{ J. Fluid Mech}, vol. 216, pp. 161--191.

\bibitem[Takeuchi, Kubota, Miki, Okuda, and Shiroyama(2003)Takeuchi, Kubota,
  Miki, Okuda, and Shiroyama]{Takeuchi:Proc:2003}
\textbf{Takeuchi, S.; Kubota, J.; Miki, Y.; Okuda, H.; Shiroyama, A.} (2003):
  \ignore.
\newblock Change and trend of molten steel flow technology in a continous
  casting mould by electromagnetic force.
\newblock   In \emph{ Proc. EPM-Conference}, Lyon, France.

\bibitem[Thess, Votyakov, and Kolesnikov(2006)Thess, Votyakov, and
  Kolesnikov]{Thess:Votyakov:Kolesnikov:2006}
\textbf{Thess, A.; Votyakov, E.; Kolesnikov, Y.} (2006): \ignore.
\newblock {L}orentz {F}orce {V}elocimetry.
\newblock \emph{ Phys. Rev. Lett.}, vol. 96, pp. 164501.

\bibitem[Vorobev, Zikanov, Davidson, and Knaepen(2005)Vorobev, Zikanov,
  Davidson, and Knaepen]{Vorobev:etal:2005}
\textbf{Vorobev, A.; Zikanov, O.; Davidson, P.; Knaepen, B.} (2005): \ignore.
\newblock Anisotropy of magnetohydrodynamic turbulence at low magnetic
  {R}eynolds number.
\newblock \emph{ Physics of Fluids}, vol. 17, pp. 125105.

\bibitem[Widlund, Zahrai, and Bark(1998)Widlund, Zahrai, and
  Bark]{Widlund:Zahrai:Bark:1998}
\textbf{Widlund, O.; Zahrai, S.; Bark, F.} (1998): \ignore.
\newblock Development of a {R}eynolds stress closure for modelling of
  homogenous {MHD} turbulence.
\newblock \emph{ Physics of Fluids}, vol. 10, pp. 1987.

\end{thebibliography}

\lastpage
\end{document}